\documentclass{aa}
\usepackage{color}

\usepackage{graphicx}
\usepackage[font={tiny,rm}]{caption}
\usepackage[varg]{txfonts}
\usepackage{draftwatermark}
\usepackage{subfig}
\SetWatermarkScale{2}
\SetWatermarkColor[gray]{0.88}
\SetWatermarkText{}
\newcommand{\kms}{km\,s$^{-1}$}
\newcommand{\kic}{KIC~2831097}
\newcommand{\halfa}{H$_{\alpha}$}

\def\t{$\pm$}

\def\cd{d$^{-1}$}

\def\tmax{T$_{\rm max}$}

\renewcommand\makeLineNumber{}
\begin{document}

\title{The elusive chase for the first RR Lyr star in a binary system: the case of
KIC\,2831097}
\author{
E.~Poretti\inst{1,2,3}
\and
J.F.~Le~Borgne\inst{4,5,3}
\and 
M.~Correa\inst{3,6}
\and
\'A.~S\'odor\inst{7}
\and
M.~Rainer\inst{1}
\and
M.~Audejean\inst{3,8}
\and 
E.~Denoux\inst{3}
\and
N.~Esseiva\inst{9}
\and
J.~Fain\`e\inst{3,6}
\and
F.~Fumagalli\inst{3,10}
\and
R.~Nav\`es\inst{6}
\and
A.~Klotz\inst{4,3}
}
\institute{
$^{1}$INAF - Osservatorio Astronomico di Brera, via E. Bianchi 46, 23807 Merate (LC), Italy 
\email{ennio.poretti@inaf.it}
\\
$^{2}$Fundaci\'on Galileo Galilei-INAF, Rambla Jos\'{e} Ana Fernandez P\'{e}rez 7, 38712 Bre\~{n}a Baja, TF, Spain\\
$^{3}$GEOS (Groupe Europ\'een d'Observations Stellaires), 23 Parc de Levesville, 28300 Bailleau l'Ev\^eque, France\\
$^{4}$Universit\'e de Toulouse, IRAP CNRS UPS, 14 Avenue Edouard Belin, 31400 Toulouse, France\\
$^{5}$LAM, Laboratoire d’Astrophysique de Marseille, 38 Rue Fr\'ed\'eric Joliot Curie, 13013 Marseille, France\\
$^{6}$Agrupaci\'o Astron\`omica de Sabadell, C/ Prat de la Riba, s/n, 08206 Sabadell, Spain\\
$^{7}$Konkoly Observatory, HUN-REN Research Centre for Astronomy and Earth Sciences, MTA Centre of Excellence, Konkoly Thege Mikl\'os \'ut 15-17, 1121 Budapest, Hungary\\
$^{8}$Observatoire de Chinon, Astronomie en Chinonais, Mairie, Place du G\'en\'eral de Gaulle, 37500, Chinon, France\\
$^{9}$Observatoire St-Martin, 31 Grande Rue, 25330 Amathay V\'esigneux, France\\
$^{10}$Osservatorio Calina, Via Nav 17, 6914 Carona, Switzerland\\
}
\date{}

\abstract
{The lack of RR Lyr stars in binary systems is an atypical fact when we compared it to  other classes
of variables. Therefore, it has become a challenge for observers to detect an RR Lyr variable in a binary system.}
{The RR Lyr variable \kic\, was one of the most promising candidates.
The phases of maximum
brightness in the {\it Kepler} photometry showed a regular variation superimposed on a parabolic trend.
These variations in the times of maximum brightness (\tmax) were interpreted as a possible light-time travel effect (LTTE) in  a wide binary and a fast
evolutionary change in the period.}
{We planned two  spectroscopic runs with the FIES instrument mounted at the Nordic Optical Telescope to
test the hypothesis of binarity. 
The observations were programmed  at the predicted quadratures of the orbit, when the two mean radial
velocities are expected to differ by about 100~\kms.  The GEOS collaboration  
complemented the spectroscopic survey  by a photometric one. We also analysed {\it Gaia} time series and intensive
TESS photometry.}
{The radial velocity curves obtained at the quadratures show the same mean radial velocity (-203~\kms), which rules the
possibility of  an LTTE out. The constant mean radial velocity, the [Fe/H] content
determined from ground-based CCD observations, and 
the astrometric parameters supplied by {\it Gaia} allow us to infer that \kic\, is a single high-velocity metal-poor
RRc star belonging to the Galactic halo. 
We revisited {\it Kepler} photometry and detected a weak Blazhko effect consisting of an oscillation of only 1.1\% of the period 
in about 50~d. We also analysed the TESS photometry of {\it Kepler}-1601, whose photometry is contaminated by \kic.
We collected  116~\tmax\, from the ground-based campaign, 103~\tmax\, from TESS, and 4 \tmax\, 
from {\it Gaia}. In total, we have 3624 times of maximum brightness. Linear ephemerides cannot fit the whole 
dataset, but only  parts of them.
The period shows a tendency to decrease in value, as if it were an evolutionary effect, but not at a constant rate.
}
{The spectroscopic and photometric campaigns performed in the framework of a professional-amateur project
failed to confirm that \kic\, belongs to a binary system: the chase remains open. As an RR Lyr variable, \kic\, shows an
intriguing evolution of the period.}
\titlerunning{The RRc variable KIC 2831097}
\authorrunning{E. Poretti et. al.}
\maketitle
\section{Introduction} \label{intro}
While binary systems are widely used to understand the physical mechanisms of many classes of
variable stars and to fix their masses and luminosities, no binary system is known in which one of the components
is an RR Lyr variable. Searching for such a system is not only
a mere observational exercise. Our knowledge on the masses of RR Lyr stars is
based on evolutionary and pulsation models. A direct measure is still lacking.
The detection of RR Lyr stars that are tied in binary systems might provide accurate masses that might
be compared to  dynamical, pulsation,
and evolutionary masses, and consequently, that might verify the hypotheses and assumptions
of the theoretical models, for instance, to 
constrain the transition phase from the red giant branch to the horizontal branch and
the late evolution of solar-mass stars.

Unfortunately, the detection of orbital motions in
the data of RR Lyr stars is a very hard hunt and results have proved elusive. Because these variables are intrinsically
faint, they are mainly observed with photometry. Times of maxima (\tmax) are
continuously obtained to study evolutionary changes in the period \citep{2007A&A...476..307L}. 
%\LEt{***A&A discourages italics for emphasis. Please change all your italics for emphasis to upright throughout***}
{\it Observed minus Calculated} (O-C) values can also be 
scrutinised to show the light-time travel effect (LTTE) that arises from the displacement of the
RR Lyr star on its orbit and the consequent change in its distance from the observer.
Many of the candidate RR Lyrae stars in binaries were proposed by studying the O-C plots
\citep{2016MNRAS.459.4360L}. TU UMa is an interesting case 
\citep{1999AJ....118.2442W}, but an
additional long-term monitoring is necessary to unambiguously confirm the suspected 20 y period
\citep{2016A&A...589A..94L}.
The case of OGLE-BLG-RRLYR-02792 was more promising:
Eclipses were observed to be superimposed on an RR~Lyr-like light curve. The star turned
out to be overluminous, however, and it was  identified as 
a binary evolution pulsator \citep[BEP; ][]{2012Natur.484...75P, 2017MNRAS.466.2842K}.
%or better as a binary evolution pulsator \citep[BEP; ][]{2017MNRAS.466.2842K}.
The search for binary RR~Lyr in the OGLE-III Galactic bulge data using the O-C method led to a final sample
of 12 firm binary candidates \citep{2015MNRAS.449L.113H}, while a search in the entire 
%\LEt{***please provide the spelled-out names of all instruments and surveys at first occurrence in the main text. This may be done as a footnote if it interrupts the flow of the sentence too much. Please check throughout and amend as required. I'll not highlight this again to avoid cluttering the ms***} 
Optical Gravitational Lensing Experiment (OGLE) database
extended the sample to 87 candidates \citep{2021ApJ...915...50H}. No candidate has an orbital period shorter than
1000~d, and the expected radial velocity amplitude is very small. Therefore, no candidates have been confirmed
spectroscopically.  

Systematic spectroscopic surveys are rare because the  RR~Lyr stars are faint, which
places the targets beyond the limit of many spectrographs. A recent comparison  of new and literature
systemic (or centre--of--mass) radial velocities of 19 RR~Lyr stars once more highlighted the intriguing case of TU UMa and 
suggested three new candidates, 
SS Leo, ST Leo, and AO Peg \citep{2021AJ....162..117B}. 
Moreover, radial velocity folded curves are needed to show the changes in the systemic velocity caused by the orbital
motion, and this implies that each target must be followed for several nights.
This type of very time-consuming survey has
just started, and several candidates have been proposed, but no clear case has been found so far  \citep{2016CoKon.105..145G}.

The  proper motion (PM) values provided by  {\it Hipparcos} and {\it Gaia}  allowed  the
investigation of PM anomalies to verify whether they can be ascribed to binarity \citep{2019A&A...623A.116K}.
Moreover, stars with a common PM might also be members of wide gravitationally bound systems
\citep{2019A&A...623A.117K}. The examination of the 24 RR~Lyr variables among the 2143 stars
with PM anomalies or common PM did not provide any sure case, but a sample of 
nine candidates.
%\LEt{***"potential" and "candidate" is a tautology. A "candidate RR Lyr variable" is a star that is potentially an RR Lyr variable***} candidates. 
As a byproduct of this investigation, 
possible slow-phase changes were found in the Blazhko star ST Pic, which might be compatible 
with an LTTE \citep{2025A&A...695L..14A}. We recall that the Blazhko effect is a periodic or quasi-periodic
variation in the amplitude and/or phase of the light curve. It was discovered by Sergey Bla\v{z}ko and 
Lidiya Petrovna Tseraskaya in 1907 \citep{1907AN....175..325B}. 

\cite{2024A&A...691A.108S} proposed a method for separating the Blazhko effect from   
LTTE or any other cause (e.g. secular period changes or binarity) and applied it
to the oscillating O-C values of  
V1109~Cas. No firm conclusion about binarity was reached, but  the possibility 
of a LTTE remains.

\section{The case of \kic}
The {\it Kepler} field includes many RR Lyr variables, both already known objects
and new discoveries
\citep{2010MNRAS.409.1585B}. The almost continuous light curve 
over 4 years supplies a testbed for the
analysis of the O-C values in search for cyclic patterns caused by the LTTE.

\kic\, has been  proposed as a very promising candidate  \citep{2017MNRAS.465L...1S}.
It is a first-overtone RR~Lyr variable (RRc in the nomenclature of variable stars)
with a period $P$=0.337~d and a light amplitude $\Delta K_p$=0.40~mag. 
 A small amplitude  mode has also been detected, with a period of 0.231~d. The period ratio 
of 0.613 makes \kic\, a member of the RR$_{0.61}$ group, which is composed of more than 1000 stars
\citep{2023MmSAI..94d..48N}. This period ratio is caused by  harmonics of non-radial modes of
degree $\ell$=8 or 9 \citep{2016CoKon.105...23D}. Additional peaks have also been found in the 
frequency range 1.5-2.5~\cd. They are probably non-radial modes, but the fundamental radial mode might
be among them \citep[Fig.~5 in ][]{2017MNRAS.465L...1S}. This richness of excited modes is 
common in RR~Lyr variables \citep[e.g. ][]{2010A&A...520A.108P}.

\object{KIC 2831097}  also shows a long-term
irregular modulation of about 47~d that resembles the Blazhko effect \citep{2012AJ....144...39L}.
The O-C curve was interpreted as the sum of  two superimposed effects: a linear decrease in the pulsation
period, resembling those observed in many RR Lyr variables \citep{2007A&A...476..307L}, and a
well-defined periodicity of 753~d with  a full amplitude of 0.04~d (i.e.  58~min).

This periodicity seemed a well-established fact because it was followed for almost two full cycles
and repeated itself regularly: 
it strongly indicated an LTTE.  As an intriguing 
consequence, the O-C amplitude implies  a mass of 8$M_{\sun}$ and then  
places the hypothetical companion among black hole candidates \citep{2017MNRAS.465L...1S}.
The black hole  possibility was also proposed for   MACHO$^{\star}$~J050918.712-695015.31, another RRc star 
with a possible LTTE effect, but it was considered to be not very probable due to high mass of
about 60~$M_{\sun}$ \citep{2004MNRAS.354..821D}. The case of \kic\, appears to be much more promising.

We considered the  spectroscopic verification of the LTTE more appropriate and decisive than  the photometric verification because the light variability is highly complex.
 In order to minimise the telescope-time investment, we 
selected the  observation epochs that corresponded to the maximum
approaching and receding velocities (i.e. the quadratures of the orbit)
as derived from the O-C curve.  By monitoring a few pulsation
cycles, we were able to measure the systemic velocities at these two epochs. 
Since the O-C curve has a full amplitude of 0.04~d, the light time  to
cover the orbital radius is 0.02~d=1728~s. At the speed of light,
this corresponds to 0.52$\cdot10^{9}$~km. By assuming  $P_{\rm LTTE}=$753~d  and 
a circular orbit (as suggested by the sine-shaped O-C curve), the orbital velocity is
50.2~km~s$^{-1}$. With opposite signs
at the quadratures, the difference between  the two systemic velocities is expected to be 
100.4~km~s$^{-1}$. 

This has to be compared to  the expected amplitude of the radial velocity (RV)
curve that is due to pulsation. %hould be  around 30~km~s$^{-1}$: 
The similar RRc variable CM~Leo ($P$=0.362~d)
shows a light amplitude of 0.50~mag in the $V$ band and   a
RV amplitude of 26.55~km~s$^{-1}$ \citep{2002MNRAS.336..841D}.
Therefore, two RV curves
of \kic\, obtained at the quadratures should span two largely disjoint ranges.  
The additional modes and the Blazhko periodicity have a much weaker effect than
the suspected LTTE on the O-C values \citep[Fig.~2 in][]{2017MNRAS.465L...1S}. Therefore,
they should not be able to significantly change the difference between the two systemic velocities.

\section {Observations and data analysis} 
\begin{table*}[h!]
\centering
\footnotesize
	\caption{Parameters of the least-squares fits of the RV$_{\alpha}$ measurements. 
}
\begin{tabular} {l c r  c  rr c rr c rr }
\hline
\noalign{\smallskip}
& & & & \multicolumn{2}{c}{August 2017}& & \multicolumn{2}{c}{June 2018} & & \multicolumn{2}{c}{All data}\\
\cline{5-6}\cline{8-9}\cline{11-12}
\multicolumn{1}{c}{Term} & & \multicolumn{1}{c}{Freq.} & &\multicolumn{1}{c}{Ampl.} &
\multicolumn{1}{c}{Phase} & &\multicolumn{1}{c}{Ampl.} & \multicolumn{1}{c}{Phase} & &\multicolumn{1}{c}{Ampl.} &
\multicolumn{1}{c}{Phase}\\
\multicolumn{1}{c}{} & & \multicolumn{1}{c}{[\cd]} & & \multicolumn{1}{c}{[\kms]} & \multicolumn{1}{c}{[0,2$\pi$]} & &
	\multicolumn{1}{c}{[\kms]} & \multicolumn{1}{c}{[0,2$\pi$]}& &
\multicolumn{1}{c}{[\kms]} & \multicolumn{1}{c}{[0,2$\pi$]}\\
\noalign{\smallskip}
\hline
\noalign{\smallskip}
     $f$ & & 2.652986 & & 19.63 & 2.56 & & 18.08 &  2.33& &  18.86 &  2.53 \\
 & & $\pm$0.000010 & & $\pm$0.67 &   $\pm$0.04 & & $\pm$0.77 & $\pm$0.04 & &$\pm$0.52 & $\pm$0.03 \\ 
\noalign{\smallskip}

     $2f$ & &  & & 3.93 & 3.10 & & 3.45 &  2.48& &  3.87 &  2.93 \\
 & &  & & $\pm$0.67 &   $\pm$0.18 & & $\pm$0.77 & $\pm$0.22 & &$\pm$0.52 & $\pm$0.13\\ 
\noalign{\smallskip}

     $3f$ & &  & & 1.86 & 4.02 & & 2.91 &  3.64& &  2.22 &  4.05 \\
 & &  & & $\pm$0.67 &   $\pm$0.36 & & $\pm$0.77 & $\pm$0.26 & &$\pm$0.52 & $\pm$0.23 \\ 
\noalign{\smallskip}
     
	$4f$ & &  & & 1.42 & 5.03 & & 0.58 &  6.14& & 0.86 &  5.22 \\
 & &  & & $\pm$0.67 &   $\pm$0.48 & & $\pm$0.77 & $\pm$1.29 & &$\pm$0.52 & $\pm$0.60 \\ 
\noalign{\smallskip}
\noalign{\smallskip}

	\multicolumn{3}{c}{T$_0$ [HJD]} &&\multicolumn{2}{c}{2457957.5151} &
& \multicolumn{2}{c}{2458271.1159}&& \multicolumn{2}{c}{2457957.5151}  \\
	\multicolumn{3}{c}{RV$_{\alpha,0}$ [\kms]} &&\multicolumn{2}{c}{$-203.22\pm$0.48} &
& \multicolumn{2}{c}{$-203.91\pm0.55$}&& \multicolumn{2}{c}{$-203.47\pm$0.37}  \\
\multicolumn{3}{c}{Residual r.m.s. [\kms]} &&\multicolumn{2}{c}{2.99} & & \multicolumn{2}{c}{2.94}&& \multicolumn{2}{c}{3.08}\\
\multicolumn{3}{c}{N}&&\multicolumn{2}{c}{41} && \multicolumn{2}{c}{33}& &\multicolumn{2}{c}{74}\\
\hline
\end{tabular}
\label{lsq}
\end{table*}

\kic\,  was observed with the Fibre-fed \'{E}chelle Spectrograph (FIES),  the  cross-dispersed
spectrograph mounted on the 2.56 m Nordic Optical Telescope of the Roque de los Muchachos Observatory 
in La Palma, covering the wavelength range of 3700$-$8300\,{\AA} 
\citep{2014AN....335...41T}. Because the target is faint and the expected RV amplitude is large, 
the spectra were taken with the lowest available resolving power ($R$=25\,000).
Observations of the first quadrature were performed on three  nights from  August 3 to 5, 2017 (14, 15,
and 12 spectra, respectively) and those of the second quadrature on three nights from June 1 to 3, 2018 
(13, 13, and 7, respectively). %(Table~\ref{logobs}).

The exposure time was set to 1800~sec, and the spectra were taken continuously throughout the night,
in an airmass interval ranging from 1.02 to 2.5.
The computed signal-to-noise ratio  (S/N) around \halfa\, ranges from 7.5 to 12.1 (median value 9.92)
in 2017 and from 8.2 to 12.8 (median value 10.71) in 2018.
The FIES ETC yielded S/N=13 as an expected value for an airmass of 1.3.
The agreement is fine because we took the spectra over a wide airmass range and we have to
consider the magnitude variability, not  the observing conditions alone.

We performed  
the RV extractions for each spectrum in real time at the telescope by using simple IRAF functions.
The \halfa\, was the only clear line visible in the spectra. At the end of the two observing runs,
we re-analysed  the spectra and tried different methods
to improve the results, but failed to measure any other helpful spectral line. This fact is not due only to
the low S/N, but mainly to the  metal--poor content of \kic. The use 
of the  least-squares deconvolution (LSD) technique \citep{1997MNRAS.291..658D} by using masks for metal-poor stars failed to show the mean profile.
Therefore, we can deliver RVs  computed by fitting a Gaussian to the \halfa\, line alone (RV$_{\alpha}$) 
with error bars of $\pm$2.0~\kms.

\begin{figure}[h!]
\includegraphics[width=9cm,angle=0]{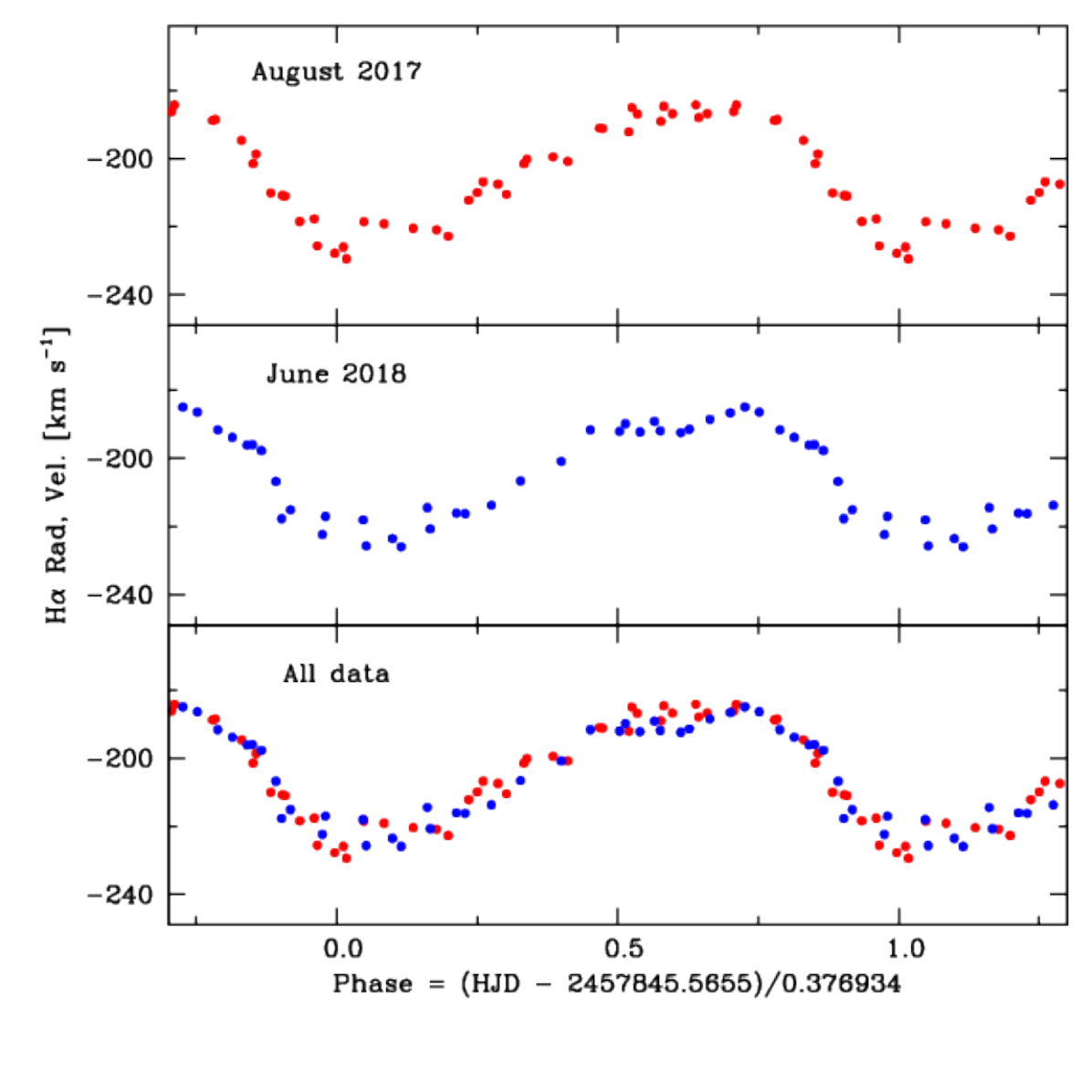}
\caption{RV measurements of \kic\, folded with the pulsation period. 
{\it Top panel:} measurements taken at the first quadrature in 2017. 
{\it Middle panel:} measurements taken at the second quadrature in 2018. 
{\it Bottom panel:} all measurements.}
\label{RVcurves}
\end{figure}

\section{The non-verification of the LTTE}
The two runs at the NOT  allowed us to answer the question about the possible LTTE
by simply folding the RV measurements with the pulsation period.
Table~\ref{lsq} reports the parameters of the least-squares fits obtained
with the formula
\begin{equation}
	RV_{\alpha}(t)= RV_{\alpha,o} + \sum_i^4 {A_i \cos [2\pi if  (t-T_o) +\phi_i ]},
\end{equation}
where $f$ is the pulsation frequency 1/P, T$_0$=HJD~2457845.5655, and $RV_{\alpha,0}$ is the average velocity
of the H$_{\alpha}$ line. 

The 41 RV$_\alpha$  measurements obtained at the first quadrature 
yield a mean velocity RV$_{\alpha,01}$=$-203.22\pm0.48$\,\kms.
The 33 RV$_\alpha$ measurements obtained at the second quadrature yield RV$_{\alpha,02}$=$-203.91\pm0.55$\,\kms\, (Table~\ref{lsq}). 
 The two mean velocities are coincident within the error bars, and the O-C variations can therefore not be ascribed to  an LTTE. We  recall that the LTTE, if real, should have
had a full amplitude of 102~\kms.  On a larger extent, 
the non-significant difference between the two mean RV$_\alpha$ strongly suggests that \kic\, is a single star.

Figure~\ref{RVcurves} shows the folded RV$_\alpha$ curves.  The Doppler variation is clearly discernible, even though we used \halfa\, alone and the low-resolution mode. Not only the two mean RV$_\alpha$ 
are coincident, but the RV points of the two runs are
almost perfectly superimposed. These two facts suggest a very weak effect of the additional modes. The only difference appears
in the standstill observed at the maximum approaching velocity, which corresponds to the
start of the expansion after the full contraction of the pulsating star.  
This standstill is covered better in the 2018 data than in those from 2017, and consequently the
Fourier parameters are slightly different (Table~\ref{lsq}). 

\begin{table*}
\centering
\footnotesize
  \caption{Participants in the  photometric campaign on \kic.}
  \label{logphot}
%  \begin{tabular}{0.95\linewidth}{lrrr}
  \begin{tabular}{lc ccc cc}
\hline
%\toprule 
\noalign{\smallskip}
	  \multicolumn{1}{c}{Observer/}&\multicolumn{1}{c}{Observatory}  &   \multicolumn{1}{c}{Telescope} & \multicolumn{1}{c}{Camera} 
	  & \multicolumn{1}{c}{Filters} & \multicolumn{1}{c}{Data points}\\ 
	  \multicolumn{1}{c}{PI/Manager}&\multicolumn{1}{c}{and/or location}  &  \multicolumn{1}{c}{} & \multicolumn{1}{c}{} 
	  & \multicolumn{1}{c}{} & \multicolumn{1}{c}{}\\ 
\hline
	  \noalign{\smallskip}
	  M. Audejean & Chinon (France) & 320 mm, f/6.0 & KAF1603 & V,R & 123,262 \\
	  M. Correa &   Sirius B, Freixinet (Spain) &   Meade 12MC, 300 mm, f/5.8 &  Moravian G2 1600 &B,V,R & 225,542,415\\
	  E. Denoux & Caussade (France) & LX200ED, 280 mm, f/6.3 &  SBIG ST-2000 XM & B,V,R &564,1799,287\\ % KAI 2020M chip
	  N. Esseiva  &Amathay V\'esigneux (France) &C11 280 mm, f/10.0   &CCD kaf 1603 ME   & B,G & 109,119\\
	  J. Fain\'e & SPAICEL, Sabadell (Spain)& Newton 200 mm, f/4.0  & ATIK 414 EX & V & 322 \\ %41$^{\circ}$33'N 02$^{\circ}$06'E
	  F. Fumagalli & Carona (Switzerland) & 300 mm, f/5.0 & Moravian G2 1600 & V,R & 962,106 \\
	  J.F. Le Borgne & Calern (France) & Tarot 250 mm, f/3.0 &  Andor DW436S   &  clean & 828 \\
	  J.F. Le Borgne & OHP (France) & IRiS 500 mm,f/8.0   &FLI Proline 4240  & g,r & 385,385 \\
	  J.F. Le Borgne & Marseille (France) &  254mm, f/4.0 &  Apogee Alta F9000  &   R &        2116 \\
	  R. Naves & Montcabrer, Cabrils (Spain) & LX200ACF, 305mm, f/9.0 &  Moravian G4 9000  & V & 42 \\
	  \'A. S\'odor & Konkoly Obs., Mt. Piszk\'estet\H{o} (Hungary) & Schmidt 600/900/1800 mm &  FLI ProLine PL16801  & clean & 2521 \\
   \hline
\end{tabular}
\end{table*} 

\section{The photometric monitoring}
The spectroscopic observations ruled out the LTTE as a cause of the  O-C variations of \kic. This still postponed
the discovery of the first RR~Lyr in a binary system. 
To search for a different cause, we planned new ground-based photometric observations and analysed data 
	collected by the {\it Kepler} \citep{2010Sci...327..977B}, {\it Gaia}  \citep{2016A&A...595A...1G}, and TESS \citep{2015JATIS...1a4003R} spacecrafts.

	\subsection{{\it Kepler}}
{\it Kepler} monitored \kic\, almost continuously from June 20, 2009 to May 8, 2013,
thus allowing the detection of strong phase variations \citep{2017MNRAS.465L...1S}.
These variations were measured over 30-d sections of the light curve. In order to
determine individual \tmax, we slightly modified the approach to the data by
fitting the nine harmonics of the pulsation frequency 
to intervals  of 3~d. This interval ensured a good coverage of the pulsation cycle,
despite the long-cadence mode (29.4~min time resolution) and allowed us to determine
3391~\tmax\, spanning 1419~d.

The resulting linear ephemeris is
\begin{equation}
\begin{array}{lrl}
{\rm T_{max} = HJD} & 2455002.5555 & + 0.37704947\,\, \cdot{\rm E}\\
               &    \pm0.0009 &\pm0.00000043  
	\label{eqkepler}
\end{array}
\end{equation}

\begin{figure}
	\includegraphics[width=8.7cm]{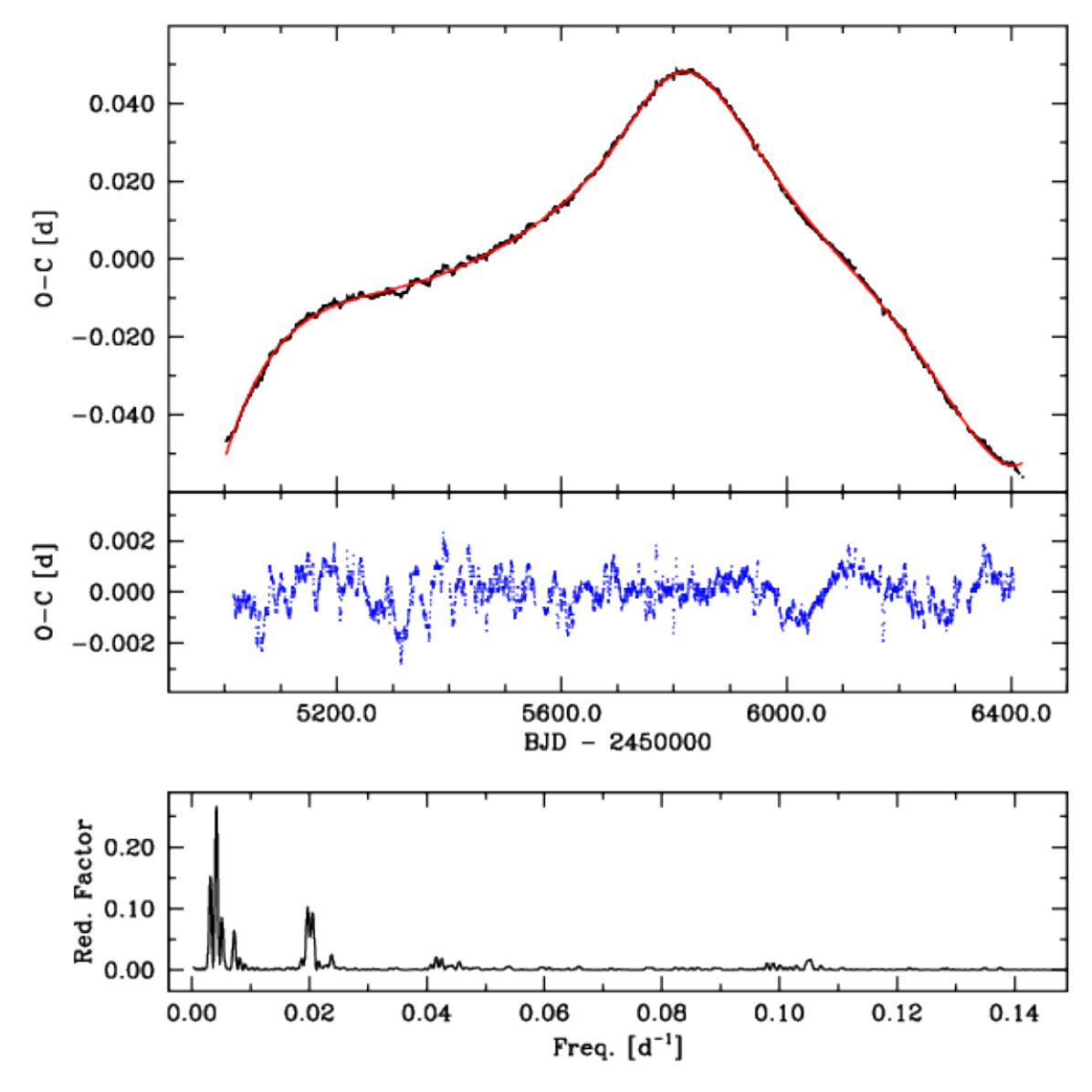}
	\caption{O-C plots of the individual maxima  observed by {\it Kepler} in 2009-13. {\it Top panel:} O-Cs 
		from the linear ephemeris fitted by cubic splines (in red). {\it Middle panel:} residual O-Cs from
		the spline fit. {\it Bottom panel:} power spectrum of the residual O-Cs.
        }
        \label{kepler}
\end{figure}

\noindent (standard deviation s.d.=0.027~d). This ephemeris  
allowed us to evaluate the change in the
period in the JD 2455002-2456420 interval (Fig.~\ref{kepler}, top panel).
Except for a small offset in the initial epoch, the shape of the O-C plot is almost identical 
to the previous one \citep[Fig.~2 in][]{2017MNRAS.465L...1S}, where  
the continuous change in the O-C values was fitted as  the 
sum of an evolutionary  decreasing period and  an orbital solution with a period of 753~d.

Because the spectroscopic campaign ruled out the binary hypothesis, we firstly fitted the whole 1419~d long O-C curve 
with a cubic spline (Fig.~\ref{kepler}, red line in the top panel) and then plotted the residual O-Cs 
(Fig.~\ref{kepler}, middle panel). 
The initial and final \tmax\, are not plotted because we know that the fit with a  cubic spline is poor at the extrema.

The residual O-Cs show some interesting features. Rapid fluctuations are clearly visible, especially in the
2455200-2455500 time interval, but they seem to be variable in shape, time and amplitude. Moreover, 
there are also slow fluctuations, as in the 2456000-2456300 time interval.  The frequency analysis
of the residual O-Cs clearly illustrates the different regimes (Fig.~\ref{kepler}, bottom panel): the 
group of peaks below 0.01~\cd\, are due to 
the slow oscillations, the double peak at 0.0187 and 0.0205~\cd\, (53.4 and 48.7~d, respectively) 
to the Blazhko effect. As it happens when a signal is not constant in phase and/or amplitude, the peak in the
power spectrum is split into two or more components. 

Our  O-C analysis confirms  the irregular nature of the Blazhko
period detected from the frequency analysis of the {\it Kepler} time series \citep[$\approx$\,47~d,][]{2017MNRAS.465L...1S}. 
We can also infer that the Blazhko effect shows a peak-to-peak amplitude of 0.004~d (Fig.~\ref{kepler}, middle panel), 
corresponding to 1.1\% of the pulsation period only.

\begin{figure*}
\includegraphics[width=13cm,angle=270]{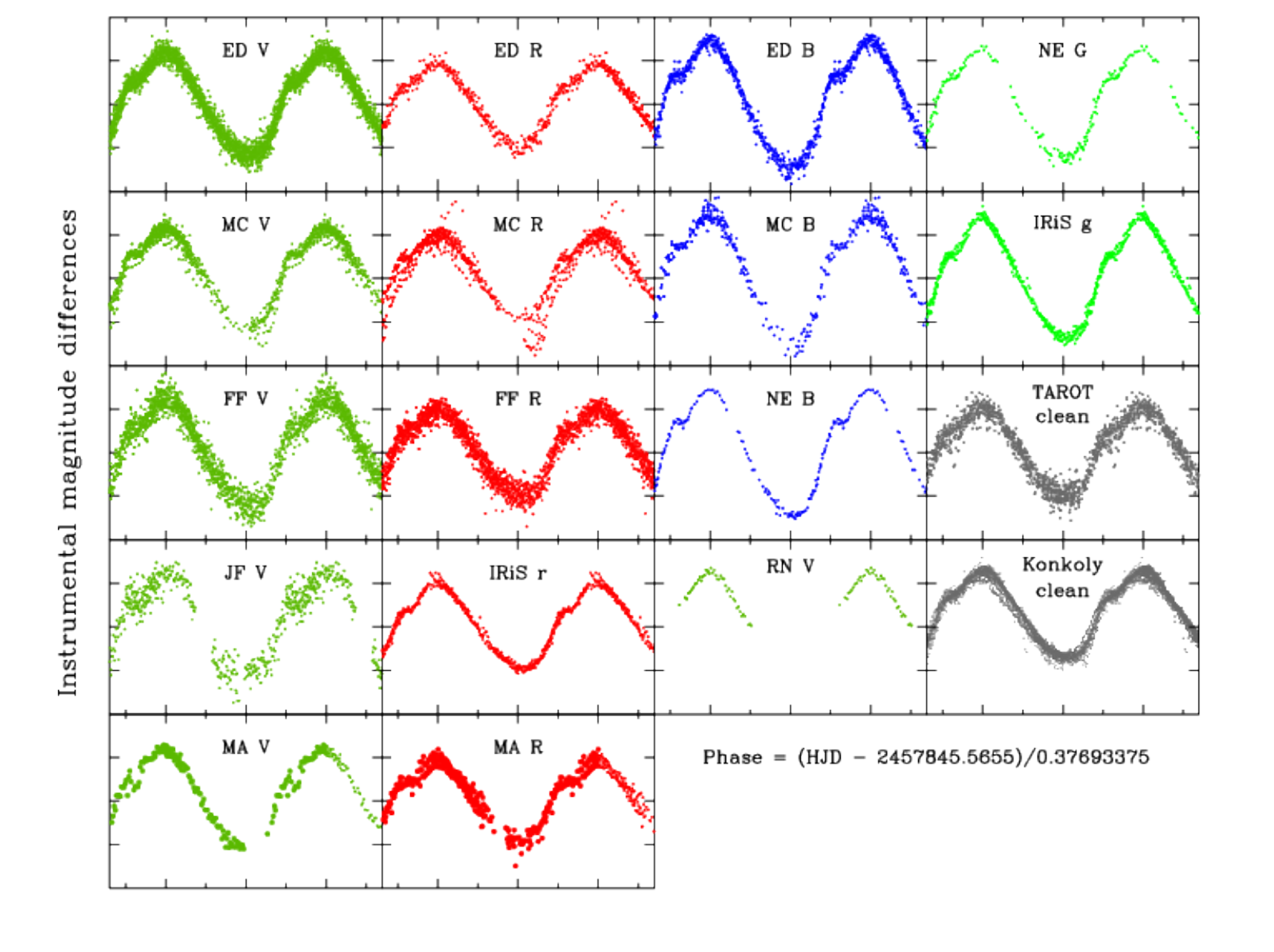}
	\caption{Light curves  of \kic\,  obtained in 2017 and 2018, folded with the pulsation period.
	The ticks on the y-axis are separated by 0.20 mag. In each panel, the initials of the
	observer or the instrument name are given together with the filter. 
%	Note the off-phase curve  in the panel ``Konkoly", obtained in 2016.
	}
	\label{merce}
\end{figure*}

\subsection{Ground-based CCD observations}
The  {\it Groupe Europ\'een d'Observations Stellaires} (GEOS) collaboration ensured a follow-up photometric coverage 
(Table~\ref{logphot}) to obtain new \tmax\, during the
spectroscopic observations. 
The campaign included telescopes located in professional astronomical
observatories (OCA, OHP and Konkoly) and in private observatories managed by well-equipped 
amateurs astronomers. Thus, the collaboration resulted 
in a very fruitful Professionals-Amateurs project.

The main goal of the simultaneous campaign was to certify that in the case of binarity
\kic\, followed the O-C periodicity that was due to the LTTE. 
Figure~\ref{merce}  shows the light curves obtained in the framework of the project. 
The mean amplitudes are 0.60~mag in $B$ light, 0.48~mag in $V$ light, and 0.40~mag in $R$ light.
The standstill before the maximum brightness appears to be more pronounced in $B$ light  than
in $R$ light.

The Fourier decomposition of the four $V$ light curves without gaps (ED, MC, MA and FF) yields 
values of $\phi_{31}=\phi_3-3\phi_1$  in the 4.67-4.98 range, with a mean value  $\phi_{31}=4.78\pm0.14$.  
With  a period of 0.377~d,  \kic\, might be a long-period RRc star (first-overtone pulsators) or a short-period RRab star
(fundamental-mode pulsators). The $\phi_{31}$
value clearly and unambiguously places it on the RRc sequence \citep[see Fig.~5 in][]{2001A&A...371..986P}.
The mean $\phi_{31}=4.78\pm0.14$ value also allowed us to estimate the metallicity by using the well-known 
$P$-[Fe/H]-$\phi_{31}$ relation for RRc stars \citep[Eq.~4 in][]{2013ApJ...773..181N}.
We obtained [Fe/H]$=-1.15\pm0.38$~dex. 

To complete the description of \kic\, as an RRc variable, we note that the RV$_{\alpha}$ amplitude of 37~\kms\, 
translates into an amplitude of the metallic lines RV$_m$=28~\kms\, 
\citep[Table~11 in ][]{2017ApJ...848...68S}. 
The RV$_m$ and $V$ full amplitudes of \kic\, (28~\kms\, and 0.48 mag) agree very well 
with the linear relation valid for RRc stars in the Galactic field and globular clusters
\citep[Fig.~8 in][]{2017ApJ...848...68S}.

All the measurements  taken in 2017 and in
2018 can be folded with the same period (Fig.~\ref{merce}) and the 111 \tmax\, can be satisfied by
the linear ephemeris (s.d.=0.0065~d)
\begin{equation}
	\begin{array}{lrl} 
{\rm T_{max}} = {\rm HJD}  & 2457845.5655 & + 0.37693375\,\, \cdot{\rm E}\\
    &\pm0.0012 &\pm0.00000139   
		\label{geos}
\end{array}
\end{equation}

\noindent that is only valid in the JD 2457845-2460820 interval.

\begin{figure}
\includegraphics[width=9.5cm,angle=270]{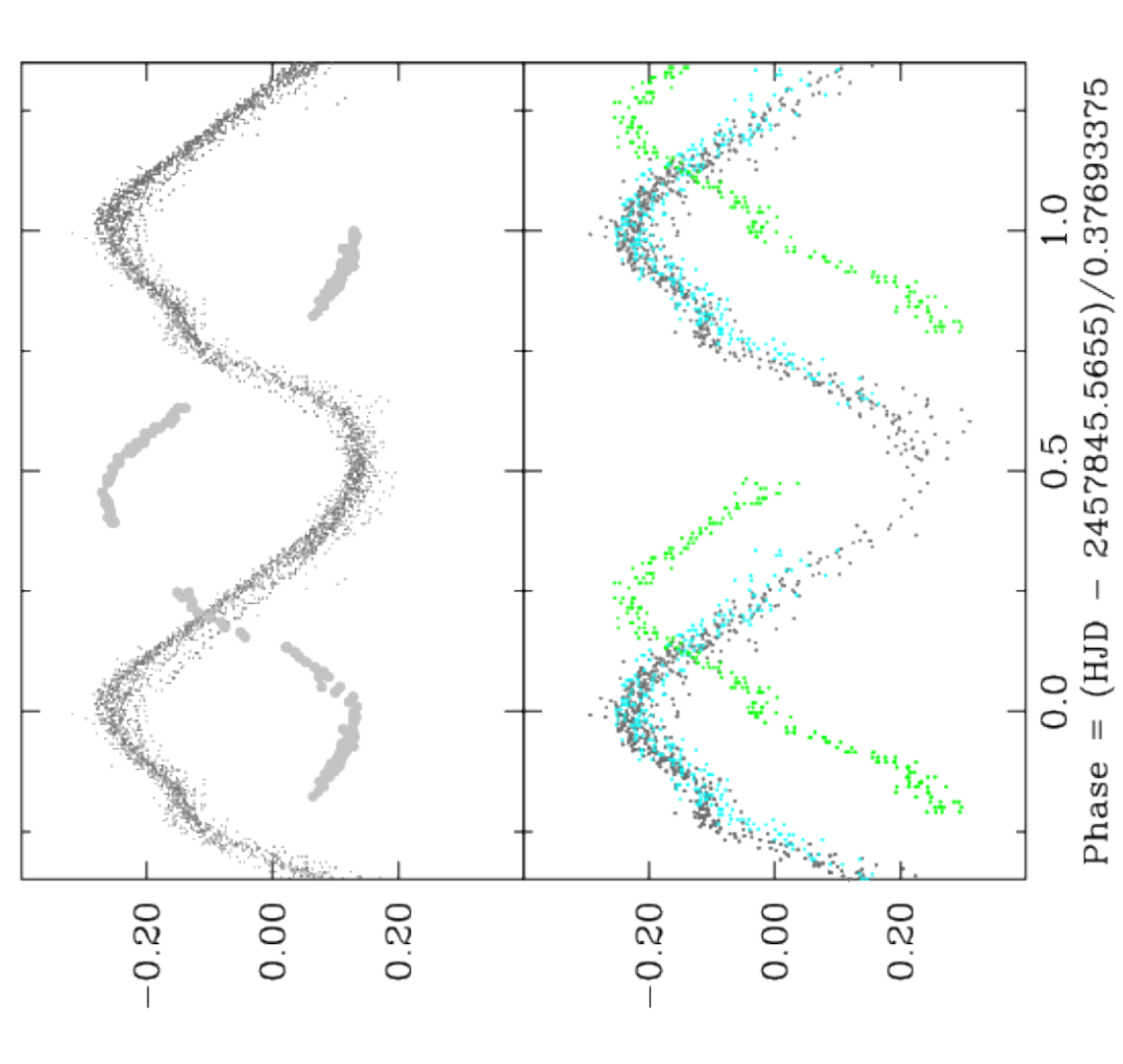}
	\caption{Large shifts of the light curve of \kic\, observed before and after the 2017-18
	campaign. {\it Upper panel:} no-filter data obtained in 2016 (light grey) and in 2017-18 (dark grey)
	at the Konkoly Observatory.
	{\it Lower panel:} $V$ light data  obtained   
 in 2017-18 (grey points), in 2020 (cyan points), and in 2022 (green points) at the Freixinet observatory.
	}
	\label{shifts}
\end{figure}

On the other hand, the measurements taken in 2016 at the Piszk\'estet\H{o}
Station of the Konkoly Observatory are completely off-phase. These measurements
are highlighted in light grey in the top panel of Fig.~\ref{shifts}.
The light curve observed in 2016
appears to be almost in phase opposition (O-C=$-0.172$~d or O-C=+0.205~d) with respect 
to those observed in 2017 and 2018 (shown in dark grey).

The time series taken in 2020 and 2022 also provide an example of a sudden shift.
The measurements obtained in 2020 (cyan points in the lower panel of Fig.~\ref{shifts}) are still in phase 
with those obtained on 2017 and 2018 (grey points): 
the three maxima yield O-Cs of +0.0033, --0.0013 and 0.0028~d.
On the other hand, the light curve obtained in 2022 (green points)
is clearly shifted forward and the observed maximum yields an O-C of +0.0886~d.

To summarise, the \tmax\, obtained by the GEOS collaboration  from 2017 to 2020 are well fitted
by a linear ephemeris (Eq.~\ref{geos}), which  is certainly no longer valid in 2016 and 2022. Therefore,
the pulsation period of \kic\, is highly variable.

\subsection{Gaia}
The {\it Gaia} satellite observed \kic\, from September 2014 to May 2017, 
collecting 47, 47 and 46 data points in the
$G$, $B_p$, and $R_p$ bands, respectively.
Table~\ref{gaia} lists the astrometric and astrophysical parameters extracted from the 
{\it Data Release 3} \citep[DR3; ][]{2023A&A...674A...1G}.
The {\it Gaia} pulsation parameters are also listed \citep{2023A&A...674A..18C}. 
For sake of completeness, 
we added our determinations of the  radial velocity because no value is reported in DR3
and of the [Fe/H] value that we obtained from our more reliable CCD light curves (Fig.~\ref{merce}).

\begin{figure}
\includegraphics[width=9.0cm,angle=0]{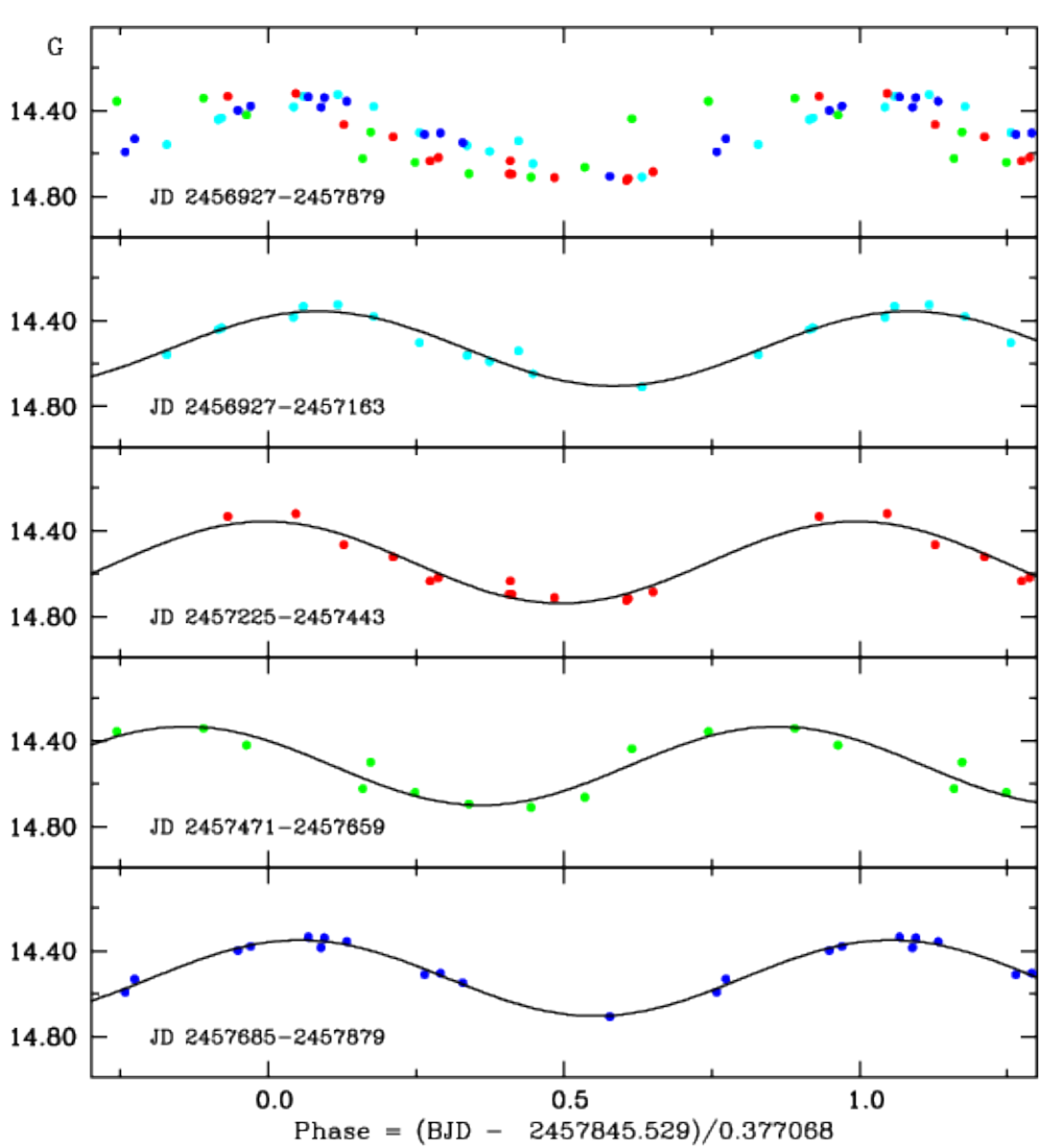}
	\caption{{\it Gaia} photometry in the $G$-band folded with the pulsation period. 
	{\it First panel from top:} all the measurements taken. 
	{\it Second panel:} measurements taken from JD 2456927 to 2457163. 
	{\it Third  panel:} measurements taken from JD 2457225 to 2457443. 
	{\it Fourth panel:} measurements taken from JD 2457471 to 2457659. 
	{\it Bottom panel:} measurements taken from JD 2457685 to 2457879.}
	\label{gaialc}
\end{figure}

Figure~\ref{gaialc} shows the
{\it Gaia} photometry in the $G$, $B_p$, and $R_p$  bands folded with the DR3 period, i.e. 0.377068$\pm$0.000008~d. 
We obtained a poorly defined light curve when folding all the measurements (top panel). This is a clear
sign that the period of \kic\, changed during the three years of the {\it Gaia} survey. Therefore, we subdivided the
dataset into four subsets, each spanning about 200~d. The scatter of each  light curve was then strongly reduced, and the 
phase shifts caused by the changing period are clearly detected (second, third, fourth, and fifth panel). 
The same results were obtained for  $R_p$ and $B_p$ photometry.
The DR3 amplitudes (0.43, 0.36, and 0.25 in $B_p$, $G$, and $R_p$, respectively)
decrease from blue to red, like those of GEOS.
%We can also measure the phase shifts by 

We computed the observed \tmax\, by interpolating the data points with a sinusoid 
(the solid line in Fig.~\ref{gaialc}). These four \tmax\, (Table~\ref{gaia}) bridge the gap between the {\it Kepler}
\tmax\, and those of the 2017-18 campaign and greatly helped us to understand 
the complicated O-C behaviour. We immediately noted that they perfectly bracket the 2016 campaign and  support
the occurrence of a large phase shift. 

{\it Gaia} DR3 reports a distance $d$=3.85$\pm$0.12~kpc, as obtained from  the  General Stellar Parametrizer 
from Photometry (GSP-Phot). $B$ and $V$ magnitudes of \kic\, are provided by  the 
Exoplanet Follow-up Observation Program (ExoFOP\footnote{\tt https://exofop.ipac.caltech.edu/tess/}), i.e. 
$B=15.465\pm0.068$ and $V=14.907\pm0.115$. 
Because RRc variables are early-type A stars, $B-V$=+0.56$\pm$0.13 is affected by a high colour excess,
%The $B-V$ colour (+0.56$\pm$0.13) is large for early A-type stars like RRc variables.  The colour excess 
as proven by the heavy interstellar absorption measured by {\it Gaia} ($A_G$=1.335$\pm$0.008~mag).
We obtained an absolute magnitude $M_V=0.64\pm0.12$ by combining distance $d$,  magnitude $V$, and absorption $A_G$,
which agrees very well with the $M_{V,RRc}=+0.59\pm0.10$ value attributed to RRc stars \citep{2013ApJ...775...57K}.

In this respect,  the {\it Gaia} DR3 trigonometric 
parallax (0.1268$\pm$0.0156~mas) seems to be underestimated and is probably affected by  a large bias 
\citep{2021A&A...649A...4L}. We also note the large difference compared to the DR2 trigonometric parallax,
i.e. 0.0394$\pm$0.0197~mas.

The Galactic coordinates, the low metallicity  and the high radial velocity strongly suggest that \kic\, is an RRc
star belonging to the Galactic halo, where RR~Lyr stars are commonly observed 
\citep[e.g. ][]{2019MNRAS.482.5327G}.

\begin{table}
\centering
\footnotesize
	  \caption {{\it Gaia} parameters of \kic, completed with our measurements of RV${_\alpha}$,
	  [Fe/H] and  \tmax$_G$.} 
	\begin{tabular}{l c c}%  rr c rr c rr}
\hline
		\noalign{\smallskip}
		\multicolumn{1}{c}{Parameter}    && \multicolumn{1}{c}{Value} \\
		\noalign{\smallskip}
		\hline
		\noalign{\smallskip}
		{\it Gaia} ID  &&  DR3~2099912216573778432 \\
		Gal. Long. $l$ [deg, 2016.0] && 68.6584372558 \\ %2016
		Gal. Lat. $b$ [deg,  2016.0] 	&& 14.3967978874	\\ %2016
		$\mu_{\rm RA}$ [mas yr$^{-1}$]  &&  $-2.6332\pm$0.0167\\
		$\mu_{\rm Dec}$ [mas yr$^{-1}$]  && $-4.4392\pm$0.0175\\
		Parallax [mas]	&& 0.1268 $\pm$ 0.0156	\\
		G [mag] && 14.5349$\pm$0.0071\\
		A$_G$ [mag] && 	1.335$\pm$0.008\\
		Distance$_{\it (GSP-Phot)}$ [pc] && 3850$\pm$120 \\
		H$_{\alpha}$ radial velocity [\kms]&&{$-203.43\pm$0.42} (\it This paper)\\
		\noalign{\smallskip}
		T$_{\rm eff}$ [K] &&  11852$\pm$120 \\ %$^{+110}_{-126}$ \\
		$\log g$ [cm~s$^{-2}$] &&  4.041$\pm$0.025\\ % $^{+0.025}_{-0.024}$ \\
		$\rm{[M/H]}$ [dex]  &&	$-0.863^{+0.215}_{-0.102}$\\
		$\rm{[Fe/H]}$ [dex]  &&	$-1.15\pm0.38$ (\it This paper)\\
		\noalign{\smallskip}
                Period [d] && 0.377068$\pm$0.000008\\
                First \tmax$_G$ [BJD] && 2457045.043$\pm$0.042 (\it This paper)\\
                Second \tmax$_G$ [BJD] && 2457334.191$\pm$0.042 (\it This paper)\\
                Third  \tmax$_G$[BJD] && 2457565.691$\pm$0.053 (\it This paper)\\
                Fourth  \tmax$_G$[BJD] && 2457782.199$\pm$0.030 (\it This paper)\\
		\noalign{\smallskip}
\hline
  \end{tabular}
	\label{gaia}
\end{table}

	\subsection{TESS}

There is no TESS light curve of \kic$\equiv$TIC~120531707  in the 
Mikulski Archive for Space Telescopes\footnote{\tt https://exo.mast.stsci.edu} (MAST).
Nevertheless, we submitted  several queries to the archive, also using {\tt lightkurve} \citep{2018ascl.soft12013L}. 
Our attention was drawn by the light curves of the nearby {\it Kepler}-1601$\equiv$\object{TIC 120531708},
1.03~mag brighter than \kic.
{\it Kepler}-1601 harbours a confirmed transiting exoplanet with an orbital period of
2.209218$\pm$0.000014~d and no significant transiting time variations \citep{2019RAA....19...41G}. No spectroscopic
follow-up is available so far.

\begin{figure}
\centering
\subfloat{\includegraphics[width=0.95\linewidth]{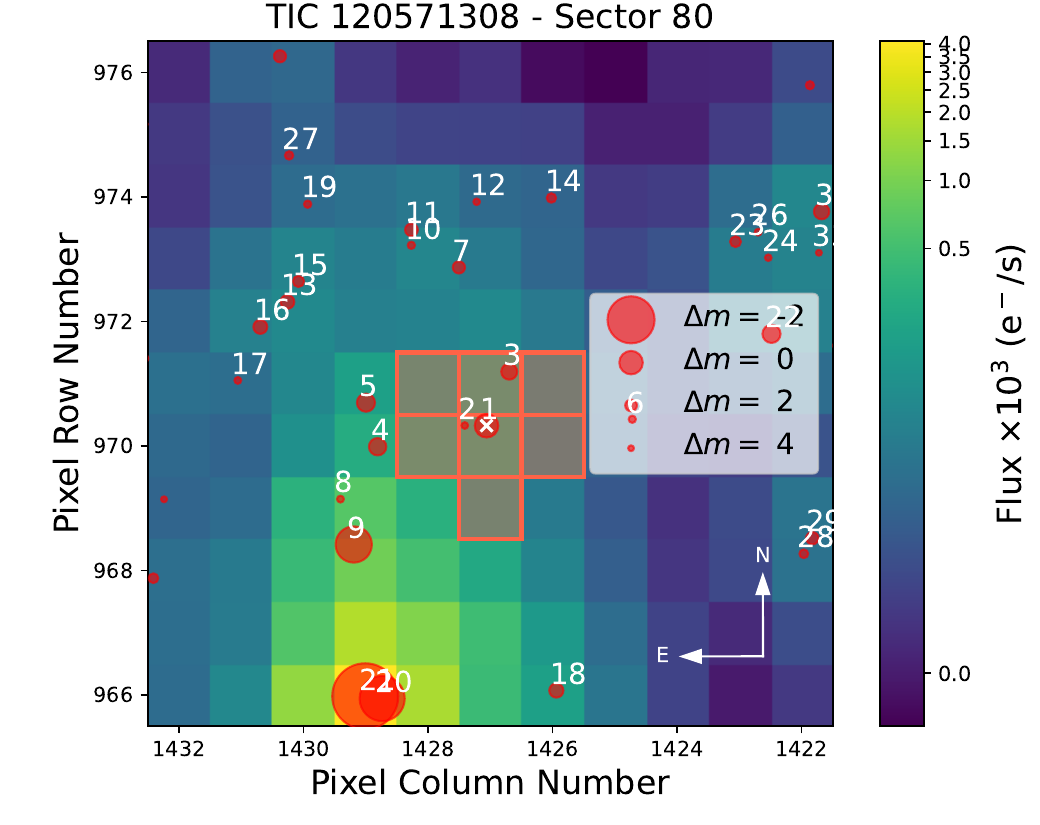}}
\\
\subfloat{\includegraphics[width=0.95\linewidth]{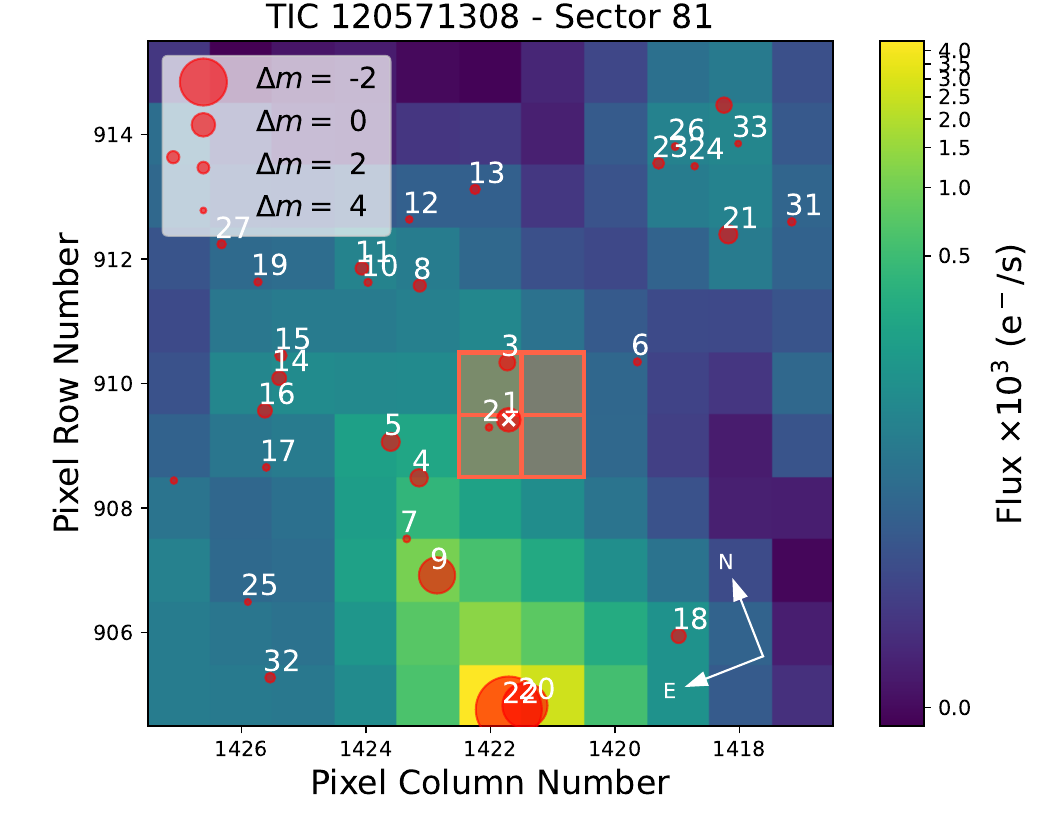}}
\caption{\captionsize{ 
TESS target pixel file of Sectors 80 ({\it upper panel}) and 81 ({\it lower}) centred on {\it Kepler}-1601. 
The red circle labelled 1 is marked by a white cross.
The orange squares delimit the pixels used by the SPOC pipeline to perform
the aperture photometry of {\it Kepler}-1601.
\kic\, is the star labelled 4 in both panels.
All sources  from the {\it Gaia} DR3 catalogue down to a magnitude difference of 4 are
shown as red circles, and the size is shown in the inserted legend.
The colour bar on the right indicates the electron counts for each
pixel.   
Both plots were  built using the code {\tt tpfplotter} \citep{2020A&A...635A.128A}. 
	}}
\label{tpfplotter}
\end{figure}

{\it Kepler}-1601 was  observed by TESS in the consecutive Sectors
80 (17276 data points)  and 81 (16958 data points) from June 17, 2024, to  August 30, 2024, with a
2-minute short-cadence.  
\kic\, is just one pixel aside of the aperture masks adopted for {\it Kepler}-1601 (Fig.~\ref{tpfplotter}).
We investigated whether this  proximity was able to contaminate the photometry of the main target.
We downloaded the time series of TIC~120531707 from the MAST, as processed by the TESS Science
Processing Operations Center \citep[SPOC;][]{2016SPIE.9913E..3EJ} at NASA Ames Research Center.    
We used the SPOC Presearch Data Conditioning Simple Aperture Photometry (PDC-SAP) light curves.

\begin{figure}
\includegraphics[width=9.0cm,angle=0]{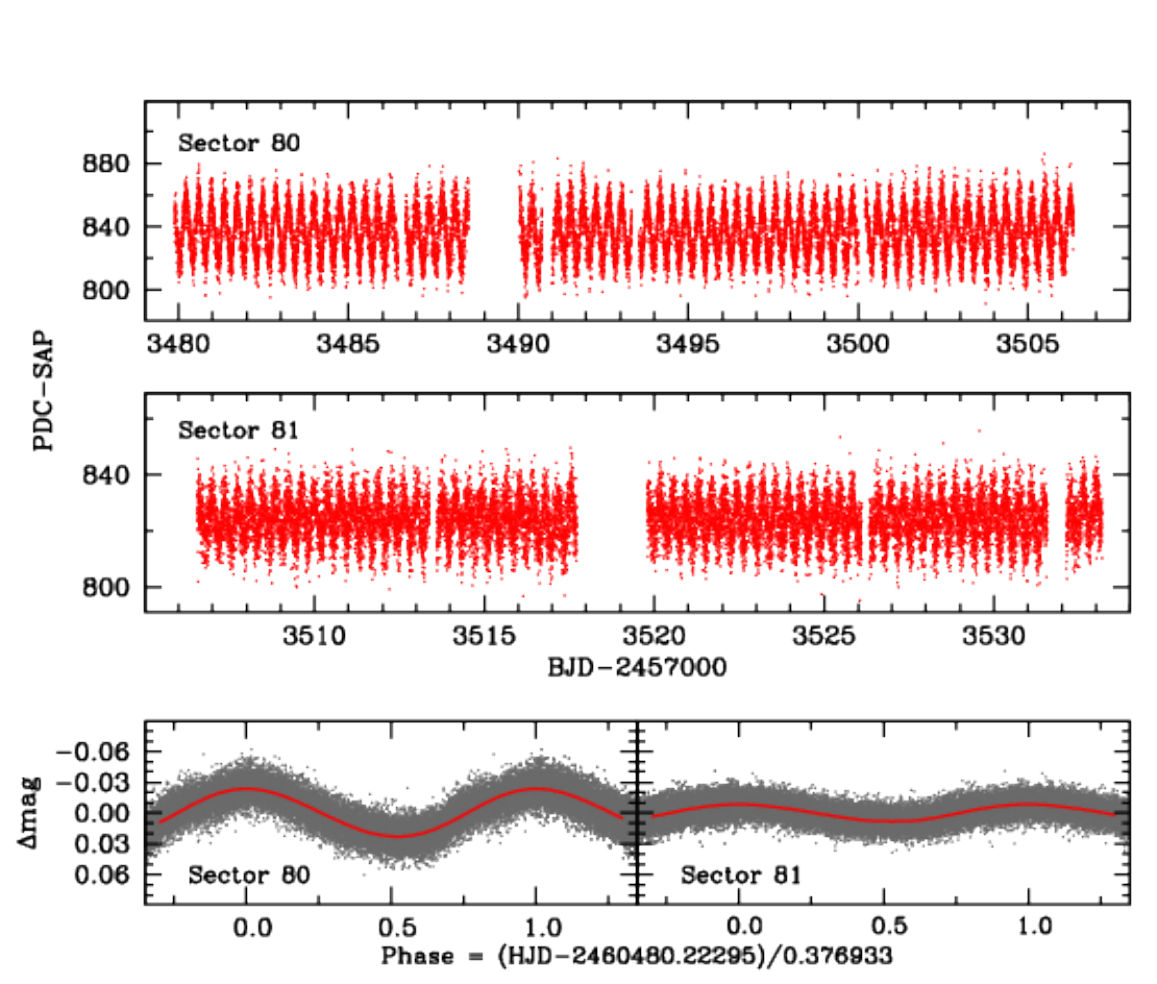}

\caption{TESS photometry of {\it Kepler}-1601 clearly showing the light
contamination due to the proximity of \kic\, to the aperture mask. The contribution
	is larger in Sector~80 ({\it top panel}) than in Sector~81  ({\it middle panel}). 
	The amplitudes of the fitting curves (in red) of the folded measurements (in grey)
	are consequently different ({\it bottom panel}).
	}
	\label{kicflu}
\end{figure}

A regular periodicity of 0.3769~d is clearly visible to eye (Fig.~\ref{kicflu}, top and middle panels), 
with a  peak-to-peak amplitude that is much larger in Sector~80 than in Sector~81.
The position of \kic\, in Sector 81 is on the bottom line of the aperture mask, and its centroid
is slightly farther away than in Sector~80 (Fig.~\ref{tpfplotter}). This reduces the amount of contaminating light.
Our interest was to study \kic, and we were therefore glad that this happened, but it  is a striking 
example of a false positive, i.e. a light variability that is not due to the main target, but to a background star. 
This subtle proximity effect can play an important role in the identification of exoplanetary candidates as well:
 an eclipsing binary located 1-2 pixels outside the adopted mask can still contaminate
the aperture photometry of the main target and mimic an exoplanetary transit.

When transformed into magnitudes, the full amplitudes of the fitting curves (the
sum of the pulsation frequency $f$  and its first harmonic
$2f$)  are damped from 0.40 mag to 0.047~mag in Sector 80 and to 0.017~mag in Sector 81 
(Fig.~\ref{kicflu}, bottom panel).
These dampings imply that the light from \kic\, contributes to the PDC-SAP flux of {\it Kepler-1601}
from 11\% (at the phase of mimimum brightness) to 16\% (at the phase of maximum brightness)
in Sector 80 and from 3.7\% to  5.4\% in Sector 81. We
note that the stars labelled 2 and 3 are even stronger  contaminants than \kic\, (Fig.~\ref{tpfplotter}). 
Our goal was to  determine the  \tmax\, of \kic\,, and we therefore did not try to improve
 the {\it Kepler-1601} photometry.

The TESS time baseline  covers 131 pulsation cycles 
and 114 were captured entirely or in part. The gaps in the light curves (Fig.~\ref{kicflu}) 
are due to satellite manoeuvres.
Eleven years after {\it Kepler} and six years after the ground-based
campaigns, the TESS survey can add significant
pieces to the puzzle of the period evolution, having in mind the
fluctuations in the O-C plot obtained from {\it Kepler} data and the off-phase
cycles observed in the GEOS campaigns. 

The standard deviations of the fits of the folded light curves are 0.0105 and 0.0083~mag in Sectors 80 and 81, respectively.
We subdivided the TESS time series into single-pulsation cycles. We computed
the least-squares fits with the $f$ and $2f$ terms and determined their 
extrema.
We were able to extract 103 reliable \tmax, disregarding those with an
unsatisfactory phase coverage or an excessive scatter. 
The resulting linear ephemeris is
\begin{equation}
	\begin{array}{lrl}
		{\rm T_{max}} = {\rm BJD}& 2460480.22295 & + 0.3769327\,\, \cdot{\rm E}\\
               &    \pm0.00114 &\pm0.0000154  
		\label{efftess}
\end{array}
\end{equation}

\noindent (s.d.=0.0060~d), which is valid in the JD 2460480-2560820 interval. 

This ephemeris disagrees with the previous one (Eq.~\ref{geos}) and suggests that a period change has
occurred. 
Since our analysis of the \tmax\, has repeatedly proven  that the period of \kic\, is rapidly variable,
we resumed GEOS observations when this paper was near completion. Four new \tmax\,
collected in May-June 2025 occurred  close to the times predicted by Eq.~\ref{efftess},
though with increasing positive O-C values.    

\section{Discussion and conclusions}

Our spectroscopic campaign demonstrated that the RR Lyr star \kic\,  is not in a binary system because the systemic 
RVs  measured with FIES are  constant within a few 0.1~\kms\, and
within the error bars. 
This result is very similar to the result obtained on Z~CVn, an  RRab star that was also suspected to be a binary member on the basis of the
O-C diagram. The binary hypothesis was later disproved by spectroscopic RV measurements 
\citep{2018MNRAS.474..824S}. Apparently, the wide variety of O-C changes in RR Lyrae stars sometimes produces behaviours
that resemble LTTE. The cases of \kic\, and Z~CVn caution us about these occurrences. 

Detailed theoretical investigations are required to determine why RR Lyr are rarely found in binary systems.
The most plausible explanation is that the evolution on the  red giant branch (RGB)
destroys most of the  potential RR~Lyr variables in binary systems \citep{2024MNRAS.52712196B}.
The core-helium burning
star underwent a mass-stripping process whose entity depends on the binary parameters,
such as orbital period and mass ratio. Complete stripping leads to subdwarf B stars,
and near-complete stripping leads to colder subdwarf A stars. Only partial stripping is able 
to produce stars that can lie on the intersection of the horizontal branch with the
instability strip and then pulsate as RR~Lyr variables \citep{2024MNRAS.52712196B}. 
The RR~Lyr subgroup of evolved RGB binaries should
have long periods (P$_{\rm orb}>1000~$d) and low mass-ratios ($0.45<q<2.0$).
The detection is therefore a hard  observation task, both in photometry and
spectroscopy, which explains the limited success achieved so far, i.e. only  lists
of long-period candidates
\citep{2015MNRAS.449L.113H, 2021ApJ...915...50H, 2016CoKon.105..145G}.

When LTTE has to be discarded as the cause,  
we have to consider the O-C variations of \kic\,  as a peculiar case of the cyclic period
changes of RR Lyr variables. We recovered the Blazhko effect in the {\it Kepler} \tmax, measuring an 
 oscillation in time of only 0.004~d.  Therefore, the O-C variations 
detected by {\it Kepler} are too large to be ascribed to this weak Blazhko effect.

\begin{table}[h!]
\centering
\footnotesize
  \caption{List of the observed times of maximum brightness. A few lines are shown for guidance, and
	the table is available in its
	 entirety at the CDS.}
  \label{listmax}
	\begin{tabular}{llll}
\hline
\noalign{\smallskip}
		\multicolumn{1}{c}{\tmax}&\multicolumn{1}{c}{Uncertainty} &  \multicolumn{1}{c}{Filter} & \multicolumn{1}{c}{Observer} \\
\hline
          \noalign{\smallskip}
		2455002.8856  &0.0003  & clean &  {\it Kepler}  \\
		         & {......} \\
		2457045.043   & 0.042      &  G &   {\it Gaia} \\
		         & {......} \\
		2457845.5629  &  0.0007    &  clean &  A. S\'odor \\
		         & {......} \\
		2457957.5145 &  0.0067    &  V &   M. Correa \\
		         & {......} \\
		2460480.217 &  0.013    & clean &   TESS \\
		         & {......} \\
		2460856.4164& 0.0062 & R & J.F. Le Borgne \\		 
   \hline
\end{tabular}
\end{table}
	We revisited {\it Kepler}, {\it Gaia}, and TESS photometry and 
observed new \tmax\, by means of a Pro-Am project that   involved several amateur 
astronomers and used  professional instruments. Table~\ref{listmax}  lists the available 3624 \tmax\,
of \kic.
The \tmax\, are also listed in the GEOS RR~Lyr database\footnote{http://rr-lyr.irap.omp.eu/dbrr/ } 
\citep{2007A&A...476..307L}, which will be continuously updated.
It is finally not   possible to find a single ephemeris that is valid for several years because  there are pieces of evidence of 
several period changes. We therefore have to limit ourselves 
to proposing linear ephemerides (Eqs.~\ref{eqkepler},\ref{geos},\ref{efftess}) that are only valid in restricted time intervals. 

\begin{figure}
\includegraphics[width=9.0cm,angle=0]{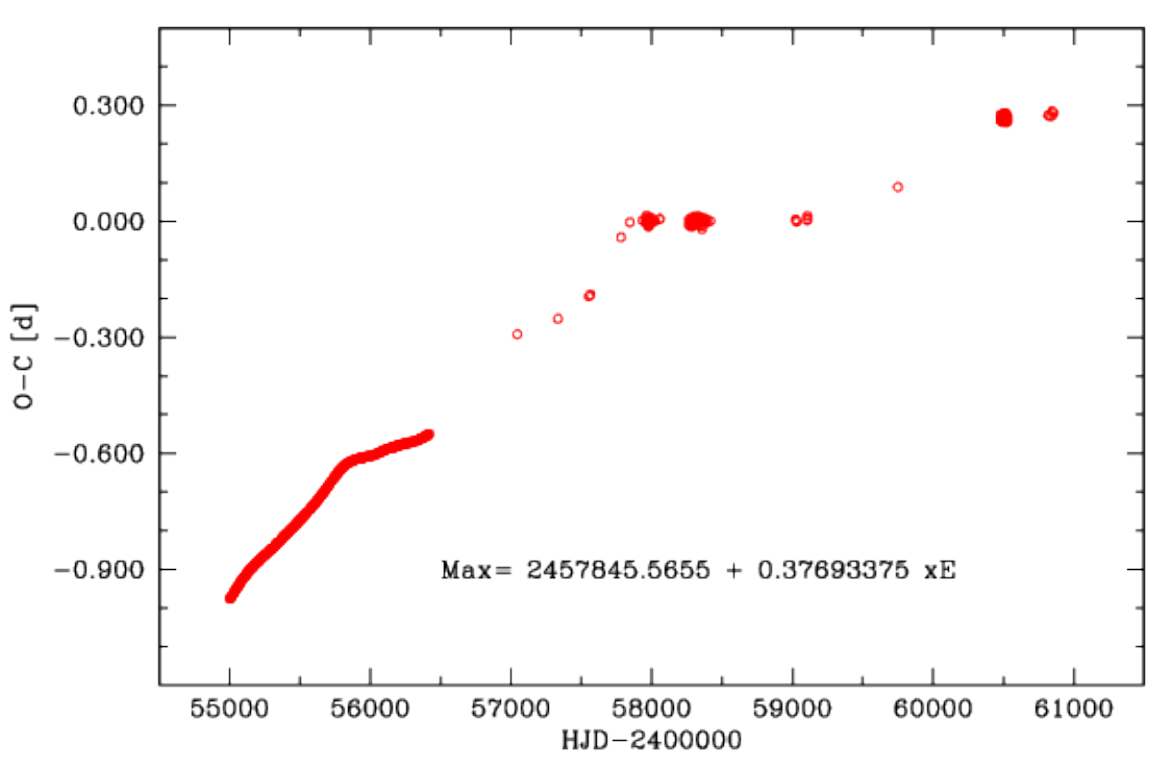}
	\caption{O-Cs of the \tmax\, computed by using a single linear ephemeris (Eq.\ref{geos}).
	}
	\label{oc6}
\end{figure}

	In order to provide the most complete picture possible, 
%we noted that the three ephemerides suggest a decreasing period and 
	we  tried to reconstruct the O-C trend 
using just one linear fit. %; we note that the three ephemerides suggest a decreasing period.} 
We emphasize that because of the gaps in the \tmax\, collection, there is a possible ambiguity 
in the number of cycles that elapsed 
between the last {\it Kepler} \tmax\, and the next one (the first {\it Gaia} \tmax). 
We were able to obtain the O-C plot shown in Fig.~\ref{oc6} by using the GEOS ephemeris (Eq.~\ref{geos}) as reference.
This plot shows the continuous decrease in the period, which is still ongoing, with superimposed  slow and 
fast oscillations. 
We must definitely conclude that the period does not seem to remain constant for a long time,
but based on the tendency to decrease in value, we might argue that it is an evolutionary
effect, if not at  a constant rate. 
Although not in a binary system, it seems that \kic\, deserves a continuous monitoring
in the next years because it might provide useful insights into the evolutionary period changes of RRc variables.

 The O-C plot of \kic\, can also be compared to those recently obtained on other RRc stars.
Differently from \kic, the periods of BE~Dor ($P$=0.328~d) and KIC~9453114 ($P$=0.366~d)
show continuous oscillations, without a clear trend to increase or decrease. 
Recently, these oscillations were tentatively ascribed to the
turbulent convection inside the hydrogen and helium ionization zones \citep{2022MNRAS.510.6050L}, which are cyclically weakened and strengthened 
owing to the presence of a transient magnetic field \citep{2006ApJ...652..643S}.
The variations in the convective envelope cannot explain the short periods of the  Blazhko effect
\citep{2012MNRAS.424...31M}, 
but they can enable the long-term ($P>100~$d) modulations.
The critical point remains the lack of pieces of  evidence that magnetic fields
are active in RR~Lyr stars, however \citep{2009A&A...498..543K}. 

On the other hand, instabilities during the final phase of the helium-core burning can
 explain the irregular period variations or the long-term oscillations,
 which are both around a constant value or superimposed on the evolutionary trend
\citep{1973ASSL...36..221S, 1979A&A....71...66S}. After the analysis of the \tmax\,
of \kic, we still consider  
evolutionary changes as the main cause of the period variations, as
proven by the agreement between theoretical predictions \citep{1991ApJ...367..524L}
and observed rate changes \citep{2007A&A...476..307L} in a large sample of RR~Lyr stars. 

\section{Data availability}
Table 4 is only available in electronic form at the CDS via anonymous ftp to cdsarc.u-strasbg.fr 
(130.79.128.5) or via http://cdsweb.u-strasbg.fr/cgi-bin/qcat?J/A+A/.

\begin{acknowledgements}
Based on observations made with the Nordic Optical Telescope, operated by the Nordic Optical 
Telescope Scientific Association at the Observatorio del Roque de los Muchachos, La Palma, Spain, under the 
proposals 55-024 and 57-014 (P.I. E.\,Poretti). E.P. thanks the whole NOT staff for the help in the observations, 
performed in visitor mode. The authors thank C\'ecil Loup and Marek Skarka for enlightening discussions.
	The outlines of this project were sketched during
several GEOS meetings, where the different knowledge of amateur
and professional astronomers found a very profitable synthesis.
The active participation of M.~Benucci, G.~Boistel, R.~Boninsegna,
R.~Dequinze, G.~Dom\`enech, M.~Dumont, J.~Fabregat, S.~Ferrand,  F.~Libotte,
J.C.~Misson, J.~Remis,  J.~Vandenbroere, J.M.~Vilalta
	to these events is gratefully acknowledged. 
\'A.S. acknowledges the financial support of the KKP-137523 ``SeismoLab" \'Elvonal grant of the Hungarian Research, Development and Innovation Office (NKFIH).
We made extensive use of the SIMBAD database, operated at CDS, Strasbourg, France.
This work has made use of data from the European Space Agency (ESA) mission
{\it Gaia} (\url{https://www.cosmos.esa.int/gaia}), processed by the {\it Gaia}
Data Processing and Analysis Consortium (DPAC,
\url{https://www.cosmos.esa.int/web/gaia/dpac/consortium}). Funding for the DPAC
has been provided by national institutions, in particular the institutions
participating in the {\it Gaia} Multilateral Agreement.
This research has made use of the {\it Exoplanet Follow-up Observation Program} (ExoFOP; DOI: 10.26134/ExoFOP5) website, which 
is operated by the California Institute of Technology, under contract with the National Aeronautics and Space 
Administration under the Exoplanet Exploration Program.
This research has made use of {\tt Lightkurve}, a Python package for Kepler and TESS data analysis 
 and 
of \texttt{tpfplotter} by J. Lillo-Box (publicly available in www.github.com/jlillo/tpfplotter);
both use in turn the python packages \texttt{astropy}, \texttt{matplotlib} and \texttt{numpy}.
\end{acknowledgements}

\bibliographystyle{aa}
\bibliography{PorettiE}
\end{document}